\documentclass[11pt,twoside]{article}

\usepackage{asp2004}
\usepackage{epsf}
\usepackage{psfig}
\usepackage{lscape}

\begin{document}
\title{Submillimeter studies of circumstellar disks in Taurus and Orion}
\author{Jonathan Williams \& Sean Andrews}
\affil{Institute for Astronomy, University of Hawaii, Honolulu, HI 96822 }

\begin{abstract}
We highlight two recent studies of circumstellar disks in the Taurus and
Orion star forming regions.
Using the JCMT and CSO, we measure disk fluxes in Taurus
over a wide range of submillimeter wavelengths and determine
the frequency dependence of the dust opacity. We find clear evidence
for a systematic change in its behavior with time, most readily explained 
by grain growth.
Using the SMA, we observed the protoplanetary disks (proplyds)
in the Orion Trapezium cluster.
The combination of high resolution, high frequency, and high
sensitivity that this instrument provides made it possible to
resolve disks from one another,
distinguish their emission from background cloud material and
surrounding ionized gas, and to detect thermal dust emission.
This allowed us to make the first mass measurements of the proplyds
and to assess their viability for planet formation.
\end{abstract}

\section{Introduction}
Circumstellar disks are an inevitable byproduct of the collapse of a
molecular core to form a star or collection of stars because of the
enormous compression of size scales and the consequent magnification
of any slight initial rotation. These disks funnel material onto the
star but they can also form planetary systems. We now know this not only
from our own Solar System but the detection of more than one hundred planets
around other stars. To help understand the mechanisms by which planets
form, however, requires basic measurements of disk properties
such as masses and sizes.  

A great deal of work has been carried out in this area of course,
including seminal papers by Beckwith et al.~(1990) on dust masses
derived from millimeter continuum fluxes, and Dutrey et al.~(1996)
on disk sizes from millimeter interferometry.
Needless to say, the BIMA interferometer has also played a
major role in disk studies (e.g., Looney, Mundy, \& Welch 2000).

Here, we present recent work on measurements of basic disk properties
in two quite different environments. Stars form mostly singly or in
binaries in Taurus, but stars in the Trapezium cluster in Orion are
much closer together and bathed in a strong ultraviolet radiation field.
We have used the three submillimeter telescopes on Mauna Kea for these
studies and have found that extending disk observations to submillimeter
wavelengths can provide important new information.
Certainly, here in Hawaii, we agree wholeheartedly that one
telescope is never enough!
A more formal writeup of this work is available in
Andrews \& Williams (2005) and Williams, Andrews, \& Wilner (2005).

\section{A submillimeter survey of Taurus disks}
Submillimeter observations probe the cool, outer regions of circumstellar disks 
where giant planets are expected to form.  Such long-wavelength data are 
required to infer the total masses of disks, and can also be used to study the 
collisional growth of dust grains into planetesimals via changes in the 
spectral dependence of the disk opacity.  Comparisons of infrared observations 
with physical models of star-disk systems have led to a sequence of 
evolutionary stages that occur before the star arrives on the main sequence 
(Lada \& Wilking 1984; Adams \& Shu 1986; Adams et al.~1987).  In the Class I 
stage, an extended circumstellar envelope is rapidly dumping material onto a 
central protostar and a massive disk.  After the envelope is dissipated, the 
system becomes a Class II object, with a disk that is actively accreting 
material onto a central, optically visible star.  In the final Class III stage, 
\emph{at least} the inner part of the disk has been evacuated, although the 
dominant mechanism for this process remains in debate (see Hollenbach et 
al.~2000).  The most interesting possibility is that the gas and dust in the 
disk have agglomerated into larger objects in a developing planetary system.  
The disk evolution scheme highlighted above is based on the shape of the 
infrared spectral energy distribution (SED), and therefore focuses on the 
inner, warmer part of the disk.  A comparable submillimeter survey could 
therefore reveal information about how disks evolve radially.

A single-dish continuum survey at 350, 450, and 850\,$\mu$m of more than 150 
disks in the Taurus star-forming clouds has recently been completed, using both 
the SCUBA array on JCMT and the SHARC-II camera on the CSO (Andrews \& Williams 
2005).  The primary goal of that program was to take advantage of the stability 
and efficiency of the SCUBA instrument to obtain an 850\,$\mu$m sample with a 
relatively uniform flux density limit of $\sim$10\,mJy (3 $\sigma$), 
corresponding roughly to a detection threshold of disks with masses greater 
than that of Jupiter ($\sim 10^{-4}$\,M$_{\odot}$).  Compared to previous 
surveys at 1.3\,mm (Beckwith et al.~1990; Osterloh \& Beckwith 1995),
the submillimeter observations are more 
sensitive by at least a factor of 5, and considerably more uniform.

\begin{figure}[!ht]
\plotfiddle{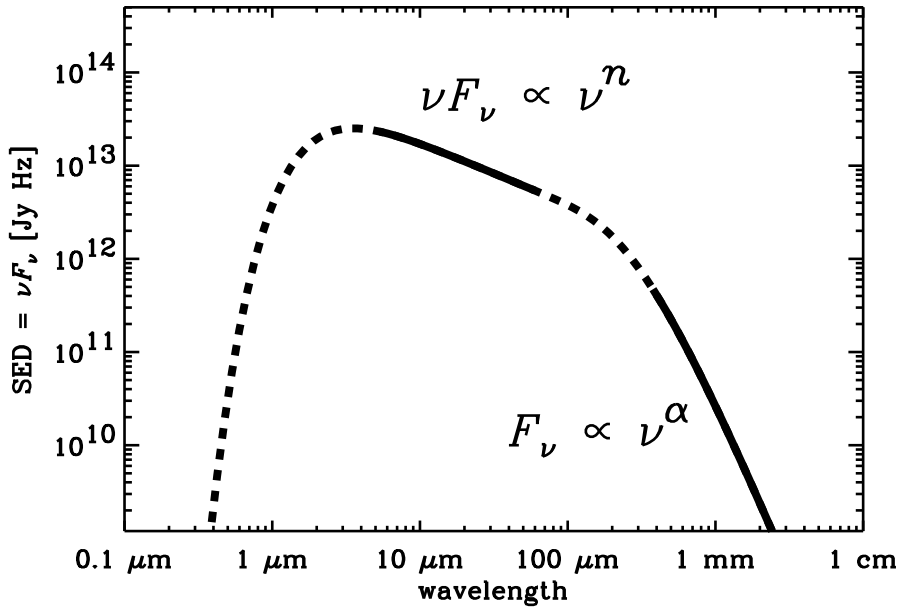}{3.5in}{0}{70}{70}{-180}{0}
\caption{A spectral energy distribution generated from a simple disk model that
assumes radial power-law distributions of both surface density and 
temperature.  The solid regions of the curve are of particular interest for disk
studies.  In the infrared, the SED behaves as $\nu F_{\nu} \propto \nu^n$, and
the index $n$ (essentially an infrared color) is a common diagnostic for
evolution in the inner disk.  In the submillimeter, the continuum spectrum
follows $F_{\nu} \propto \nu^{\alpha}$, where the index $\alpha$ is set
primarily by the spectral form of the disk opacity.}
\end{figure}

To allow for optical depth effects and a radial temperature 
distribution in determining a disk mass, a simple disk structural model (e.g., 
Adams et al.~1987; Beckwith et al.~1990) was employed in fits to the full SED 
from the infrared through millimeter.  An empirical calibration of the 
850\,$\mu$m flux density and the disk mass indicates that assuming optically 
thin emission at a temperature of $\sim$20\,K is a reasonable means of 
measuring $M_d$ in the absence of sufficient SED information.  Figure 1 shows 
an example of a SED generated from such a model, with two key regions marked: 
the mid-infrared and millimeter both are simple power laws in frequency.  
The infrared index, $n$, defines the different Classes of disk evolution
(Greene et al. 1994), and the millimeter index, $\alpha$, constrains the
dust grain opacity and thereby the maximum size of the grains
(Pollack et al. 1994).

We were able to obtain short wavelength (350 and/or 450\,$\mu$m)
fluxes for all Class I objects and more than half of the Class II
objects that were detected at 850\,$\mu$m. Combining with tabulated
fluxes at millimeter wavelengths from the literature where possible,
we then measured the SED slope over a factor of $2-10$ in wavelength range.
This allowed us to determine $\alpha$ more accurately than previous work
and study its evolution.

\begin{figure}[!ht]
\plotfiddle{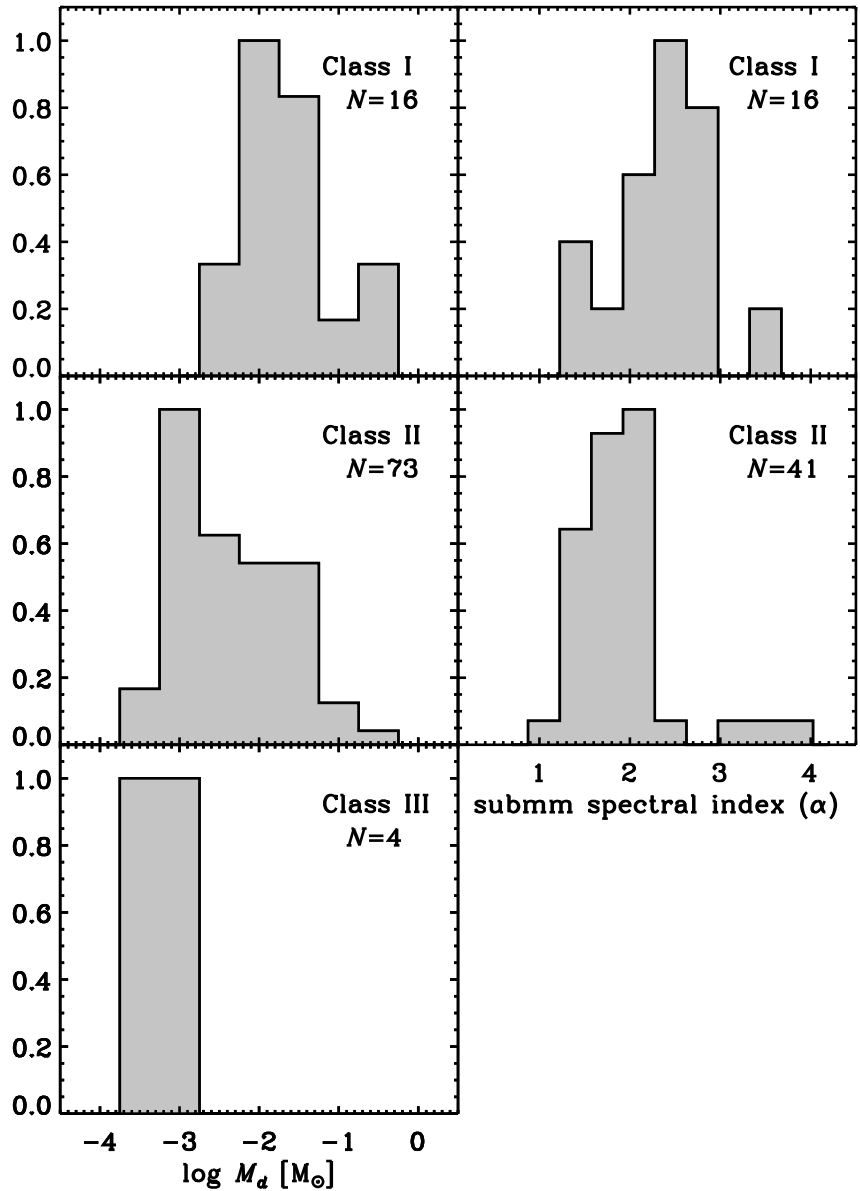}{6.2in}{0}{70}{70}{-180}{0}
\caption{Histograms of circumstellar disk masses and submillimeter spectral
indexes for different evolutionary states, derived from the JCMT-CSO survey
by Andrews \& Williams (2005). Each histogram has been normalized to a peak
of 1 to emphasize the decrease in mass from Class I to II to III, and the
decrease in $\alpha$ from Class I to II.}
\end{figure}

The major results of our survey are summarized in Figure 2.
The mass in millimeter sized grains decreases from median values of
$3\times 10^{-2}\,M_\odot$ for Class I to
$3\times 10^{-3}\,M_\odot$ for Class II to
$6\times 10^{-4}\,M_\odot$ for Class III objects.
The empirically measured millimeter spectral index also decreases
from a median of 2.5 for Class I to 1.8 for Class II objects.
This is the first time that the evolution of basic properties of
the outer disk have been quantified and is due to the high sensitivity,
large sample size, and extended wavelength coverage of the survey.

The millimeter spectral indices in Class I and Class II objects
are considerably less than that found in the ISM, $\alpha_{\rm ISM}=4$.
The low values could be caused by high optical depth or a change in
the frequency dependence of the dust grain opacity, $\kappa\propto\nu^\beta$.
A simple radiative transfer calculation shows, in the Rayleigh-Jeans limit,
$\alpha=2$ for optically thick emission and $\alpha=2+\beta$
in the optically thin case.
(A value lower than 2, as found for Class II objects, may be due to
the Rayleigh-Jeans assumption being invalid for the short wavelength data
or additional processes such as significant ionized gas emission at long
wavelengths from a jet.)
Our disk models, which integrate over the temperature and surface density
radial profiles, indicate that no more than $\sim 20-40$\% of the emission
is optically thick at submillimeter
wavelengths. We therefore conclude that $\beta$ for disk grains is not only
substantially lower than for ISM grains
(first noted by Beckwith \& Sargent 1991)
but that it also decreases as disks evolve from Class I to Class II.

Further evidence for an evolution of grain properties is shown in Figure 3.
We found an inverse correlation between the infrared and millimeter spectral
indices that can seemingly only be explained by a decrease in $\beta$.
If the grain properties did not change, $\alpha$ would depend only on
the disk structure. Typically assumed evolutionary scenarios such
as viscous spreading or decreasing disk mass would lower the optical depth
at all radii and increase $\alpha$, opposite to the observed trend.
The empirical decline in millimeter spectral slope, $\alpha$,
as disks evolve is due to a flattening of the dust opacity index,
$\beta$, and is most readily explained by an increase in the maximum
size of the grains (Pollack et al. 1994), from microns to millimeters.
(See also the paper by David Wilner in this volume for evidence of
growth to centimeter sizes in TW Hya.)

\begin{figure}[!ht]
\plotfiddle{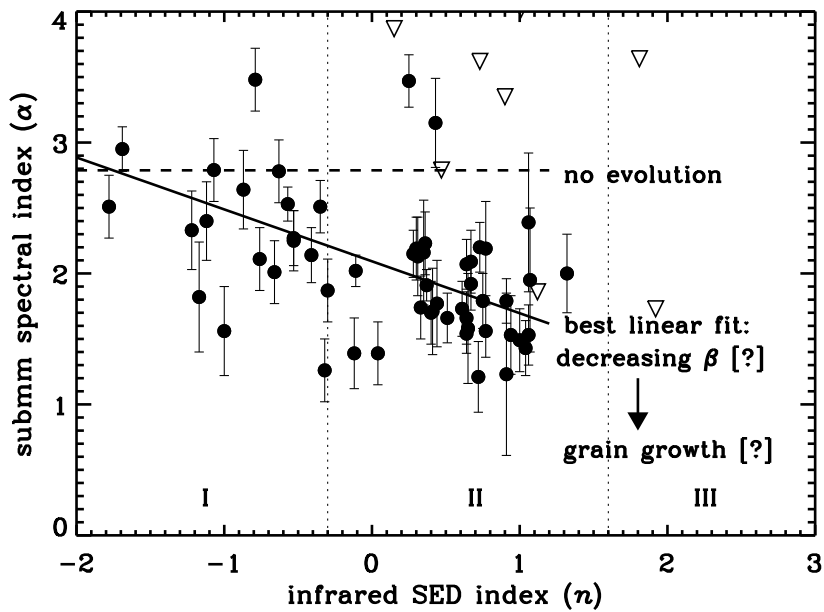}{3.5in}{0}{70}{70}{-180}{0}
\caption{The behavior of the submillimeter spectral index, $\alpha$,
as a function of the infrared SED index, $n$. The vertical dotted
lines show the boundaries between different disk classifications
based on Greene et al. (1994). Open triangles are 3 $\sigma$ upper limits.
The dashed horizontal curve marks the expected behavior if basic disk
properties do not evolve with time.
Typical evolutionary behavior, like the viscous spreading of a disk or
the loss of mass, leads to a decrease in the optical depth of the disk,
and would therefore result in a trend in the opposite sense.
The observed decrease of $\alpha$
as the disk evolves is most readily explained by grain growth resulting in
a shallower frequency dependence of dust opacity.}
\end{figure}

Understanding the timescales involved in the evolution of circumstellar disks 
is critical for placing constraints on the dominant mechanism(s) of planet 
formation.  In a purely statistical sense, our large, sensitive, and uniform 
submillimeter survey of Taurus disks enables a direct comparison with previous 
work at shorter wavelengths (e.g., Kenyon \& Hartmann 1995) to examine the 
evolutionary timescale as a function of disk radius.  We found, in general 
agreement with work at other wavelengths (e.g., Duvert et al.~2000),
that only a small fraction of disks, $<$10\%, that have no inner 
disk signatures of either infrared excess or H$\alpha$ emission
were detected in the submillimeter.  
Based on the disk frequency in clusters of different ages,
Haisch et al. (2001) estimate a lifetime for the inner disk of
$\sim 5-10$\,Myr. The outer disk must therefore become undectable
no later than a few $10^5$\,yr after the inner disk disappears.
Understanding the trigger for this rapid transition remains a key 
problem in disk evolution.  Some possible explanations for the essentially 
radially independent disk dissipation timescale include viscous accretion with 
simultaneous photoevaporation by the central star (e.g., Clarke et al.~2001),
rapid grain growth at all radii in the early stages of planet formation
(e.g., Weidenschilling \& Cuzzi 1993), or (more speculative but most
intriguing) dynamical clearing by migrating proto-planets.

\section{Submillimeter interferometry of the Orion proplyds}
Given that most low mass stars are born in OB associations
(McKee \& Williams 1997), it is essential to understand their
formation in such an environment.
As the statistics of extrasolar planetary systems become
better understood (e.g., Marcy, Cochran, \& Mayor 2000), it is also
natural to extend this question to the formation of planets
around low mass stars in massive star forming regions.

The Trapezium cluster in Orion is the nearest young, massive star
forming region and it is consequently the most intensively studied
(e.g., O'Dell 2001). There are approximatly $10^3$ $\sim 1$~Myr old stars
in the central 1~pc of the cluster core (Hillenbrand 1997) but the
radiation field is dominated by one O6 star, $\theta^1$ Ori C.
Ionized gas from the evaporating envelopes and disks around nearby
low mass stars can be observed at centimeter wavelengths with the
VLA (Churchwell et al. 1987) through the optical, most spectacularly
with HST (beginning with O'Dell, Wen \& Hu 1993).

\begin{figure}[!ht]
\plotfiddle{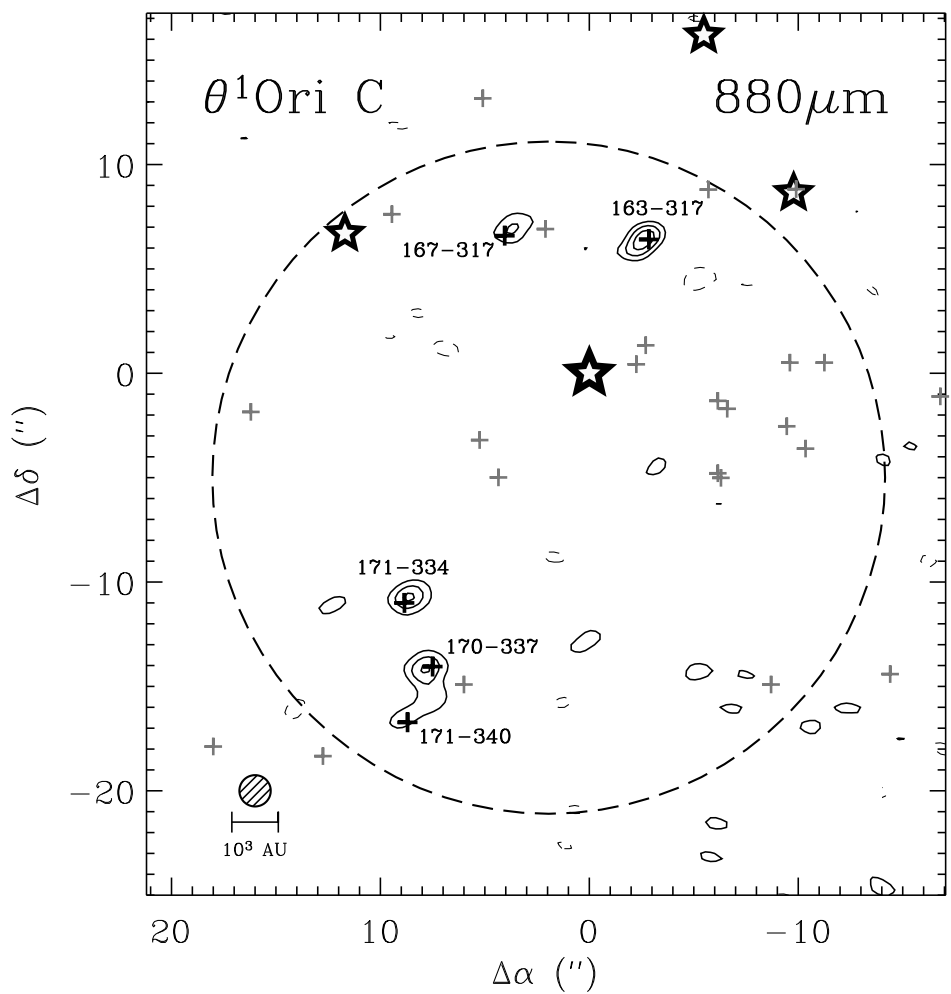}{5.3in}{0}{70}{70}{-210}{-120}
\caption{Contours of $880~\mu$m continuum emission toward the proplyds
in the Trapezium cluster. The locations of the proplyds
from O'Dell \& Wen (1994) are shown by crosses and the five
detections are labeled following their nomenclature.
The position of the four Trapezium O stars are shown by the
large star symbols and the center of the coordinate grid has
been set to $\theta^1$ Ori C. The $1.5''$ synthesized beam and scale bar
are shown in the lower left corner, the large dashed circle is
the FWHM of the primary beam. Contour levels are at
$3,5,7\times\sigma$ where $\sigma=2.7$~mJy~beam$^{-1}$
is the rms noise level in the map.}
\end{figure}

The HST images provide some of the most dramatic images of
protostellar disks that exist (see in particular Bally et al.~1998a).
They were dubbed ``proplyds'' by O'Dell as they were presumed
to be protoplanetary on account of their solar system scale sizes.
However their masses -- and potential for forming planets --
were unknown. Only a lower limit less than a Jupiter mass could be
obtained by integrating a minimal extinction over their area
(McCaughrean et al. 1998) and it was not clear,
therefore, that the proplyds had enough mass to form (giant) planets.

Disk masses are best measured
at longer wavelengths where the dust emission becomes optically thin.
Interferometry is essential to resolve the tightly clustered proplyds
from each other and also to filter out the strong emission from the
background molecular cloud.
Mundy, Looney, \& Lada (1995) used the BIMA interferometer at
$\lambda 3.5$~mm to image a field around $\theta^1$ Ori C containing 33 
proplyds.
Several significant peaks were found, four coincident with proplyds,
but the intensity was consistent with free-free emission from
ionized gas and they were unable to measure masses.
By analyzing the non-detections, however, they were able to place
a statistical upper limit of $0.03~M_\odot$ on the average disk mass.
The dust emission increases at shorter wavelengths and the free-free
emission decreases. Using the OVRO array, Bally et al.~(1998b)
imaged two fields containing a total of six proplyds at $\lambda 1.3$~mm,
made a tentative detection of one object with a mass equal to $0.02~M_\odot$,
and placed upper limits of $0.015~M_\odot$ on the other objects.
Lada (1999) presented a mosaic of two fields at $\lambda 1.3$~mm
with the Plateau de Bure interferometer that claimed three detections.
The implied masses were $\sim 0.01~M_\odot$ but these have not been
analyzed in detail.

By operating at shorter wavelengths than the other interferometers,
the Submillimeter Array (SMA; Ho et al.~2004) is better suited to
measuring the dust emission above the strong bremsstrahlung emission
from the ionized gas. Furthermore, it has a relatively large field
of view which allows many proplyds to be imaged simultaneously.

Details of the observations and their analysis are in
Williams, Andrews, \& Wilner (2005). We observed a single field
toward the center of the Trapezium cluster at 340~GHz ($880~\mu$m).
23 proplyds were contained within the $32''$ full width half maximum
primary beam. The resolution of these compact configuration data was
$1.5''$ and the rms noise level was $\sigma=2.7$~mJy~beam$^{-1}$.

Contours of the continuum emission are shown in Figure~4.
Five proplyds were detected within the $32''$ FWHM of the primary beam
with a peak flux greater than $3\sigma = 8.1$~mJy~beam$^{-1}$
and are labeled in the Figure. 167--317 ($\theta^1$\,Ori\,G)
is a very bright source in the optical and radio.
Based on an extrapolation of its SED at centimeter wavelengths
(Garay, Moran, \& Reid 1987), we appear to be detecting the
bremsstrahlung emission from its ionized cocoon even at these
short wavelengths. The four other proplyds have fluxes that are
significantly above the bremsstrahlung extrapolation and we attribute
the bulk of the SMA flux to thermal dust emission.

The corresponding disk masses for these four proplyds,
after correcting for a contribution from ionized gas,
range from 1.3 to $2.4\times 10^{-2}~M_\odot$.
These are similar to the minimum mass solar nebula (Weidenschilling 1977).
Photo-evaporative mass loss rates are high, $\sim 10^{-7}~M_\odot$~yr$^{-1}$
(Churchwell et al. 1987), but concentrated in the outer parts
of the disks where the gravitational potential of the central star
is weakest and photoevaporation is most effective
(Hollenbach et al.~2000).
The radius of the bound inner region depends on the stellar mass and
whether the gas is ionized by EUV photons or remains neutral and
only heated by the FUV radiation field. The detected proplyds lie
far enough away from $\theta^1$ Ori C for the second condition to apply and
the central $\sim 20-50$~AU radius of the disks survive (Johnstone et al.~1998).
Disk radii, measured from the HST observations, are $\sim 40$~AU for
the detections so, at most, only the outer 50\% of the disk will be lost.
For a surface density $\Sigma\sim r^{-3/2}$, the surviving mass fraction
is at least 60\%. In these systems at least, the submillimeter emission
indicates there is sufficient material bound to the star to form
Solar System scale planetary systems.

There are intriguing possibilities for the non-detections too.
We found a statistically positive flux toward the 18 proplyds
that lay below our $3\sigma$ detection limit.
The corresponding (gas + dust) mass limit is $8\times 10^{-4}~M_\odot$
and the dust-only mass limit is therefore
$8\times 10^{-6}~M_\odot\simeq 3~M_\oplus$.
That is, even if all the gas were lost from these disks,
there would still be enough material to form terrestrial-like planets.
Far more sensitive observations would be required to verify
this on an individual basis, of course, and it would also be
important to average the VLA data in a similar way to measure the
low level bremsstrahlung emission.

\clearpage
\acknowledgements
~

\noindent
JPW thanks the organizers for their invitation to this celebration of
Jack's extraordinary career. As this meeting showed, Jack's management
of the Radio Astronomy Lab allowed many students in many different areas
to flourish. I consider myself very fortunate to have been part of his
group and to have benefited from his work and his guidance.
Mahalo nui loa!

\end{document}